\documentclass[pra,aps,showpacs,twocolumn]{revtex4}
\usepackage{epsf}
\usepackage[dvips]{color}
\newcommand {\be}{\begin{equation}}
\newcommand {\ee}{\end{equation}}
\newcommand {\bea}{\begin{eqnarray}}
\newcommand {\eea}{\end{eqnarray}}

\begin{document}

\title{ Limits of sympathetic cooling of fermions:\\ The role of the heat capacity of the coolant}
\author{L.~D. Carr$^*$ and Y. Castin\\}
\affiliation{Laboratoire Kastler Brossel, Ecole Normale
Sup\'erieure, 24 rue Lhomond, 75231 Paris CEDEX 05, France}
\date{\today}

\begin{abstract}
The sympathetic cooling of an initially degenerate Fermi gas by
either an ideal  Bose gas below $T_c$ or an ideal Boltzmann gas is investigated.
It is shown that the efficiency of cooling by a  Bose gas below $T_c$
is by no means reduced when its heat capacity becomes much less than
that of the Fermi gas, where efficiency is measured by the decrease in the 
temperature of the Fermi gas per number of particles evaporated from 
the coolant.  This contradicts the intuitive idea that
an efficient coolant must have a large heat capacity.  In
contrast, for a Boltzmann gas a minimal value of the ratio of the
heat capacities is indeed necessary to achieve $T=0$ and all of
the particles must be evaporated.
\end{abstract}


\maketitle

\section{Introduction}
\label{sec:intro}

 Sympathetic cooling of fermionic atoms has proven to be 
an efficient tool to produce strongly
degenerate Fermi gases ~\cite{truscott1,schreck1,roati2002,hadzibabic2002,hadzibabic2003}.
It is often thought that when the specific heat
of the coolant becomes less than that of the fermions the
cooling process slows drastically and even comes to a
halt~\cite{truscott1,onofrio1,presilla2003}.  This belief derives from the following 
argument based on standard thermodynamics~\cite{huang1}.
A coolant is not only
described by its temperature but also by its heat capacity. The
equilibrium temperature between two bodies initially at different
temperatures $T_f$ and $T_b$ is $T=(C_f T_f+C_b T_b)/(C_f+C_b)$,
where $C_f$ and $C_b$ are the heat capacities of the two
bodies~\cite{constance}. Therefore, regardless of having a very cold object $b$
($T_b \ll T_f$), it will not act as a good coolant unless it is
also true that $C_b \gg C_f$.

However, this argument
does not include the evaporation of atoms from the coolant~\cite{hess1986,luiten1996}, 
a key ingredient of sympathetic cooling. In the following, we shall
consider the ideal case of sympathetic cooling, where all
losses of particles  due to inelastic collisions are absent. One can then evaporate sufficiently
slowly that the Fermi gas and the coolant remain close to thermal equilibrium. This
is in the spirit of the above argument, based on heat capacities,
as heat capacity is a property of an equilibrium system.
In this case, the only limitation of resources is the number of
particles in the coolant.   We then introduce an idealized model of
sympathetic cooling: the atomic motion is treated in the semi-classical
approximation and cooling proceeds in discrete steps.  In each step, the 
atoms of the coolant go through an evaporation phase, and then the fermions and
the coolant thermalize (see Fig.~\ref{fig:1} below). 
 This is a generalization of the discrete scheme 
used for evaporative cooling in Ref.~\cite{Mewes}, to the case of sympathetic cooling.
 A similar semi-classical and
discrete model for sympathetic cooling was put forward in Ref.~\cite{wouters2002}, 
with the addition of 
features like collisional losses and space
dependent effects in the evaporation process, but without a determination of the 
limits of the cooling process.  Using our idealized model, 
we shall then show that the argument based on the heat capacities 
is approximately correct for fermions cooled by
an ideal Boltzmann gas \cite{prudence}, but incorrect when cooled by an 
ideal  Bose gas below the Bose condensation temperature $T_c$. 
In fact, in the latter case, it is ~\emph{preferential} to have $C_b
\ll C_f$ to minimize the number of particles of the coolant
whose evaporation is required to reach zero temperature  \cite{proviso}.

\section{The Model}
\label{sec:model}

%%%%%%%%%%% figure 1 %%%%%%%%%%%
%
\begin{figure}[t]
\begin{center}
\epsfxsize=7.8cm \leavevmode \epsfbox{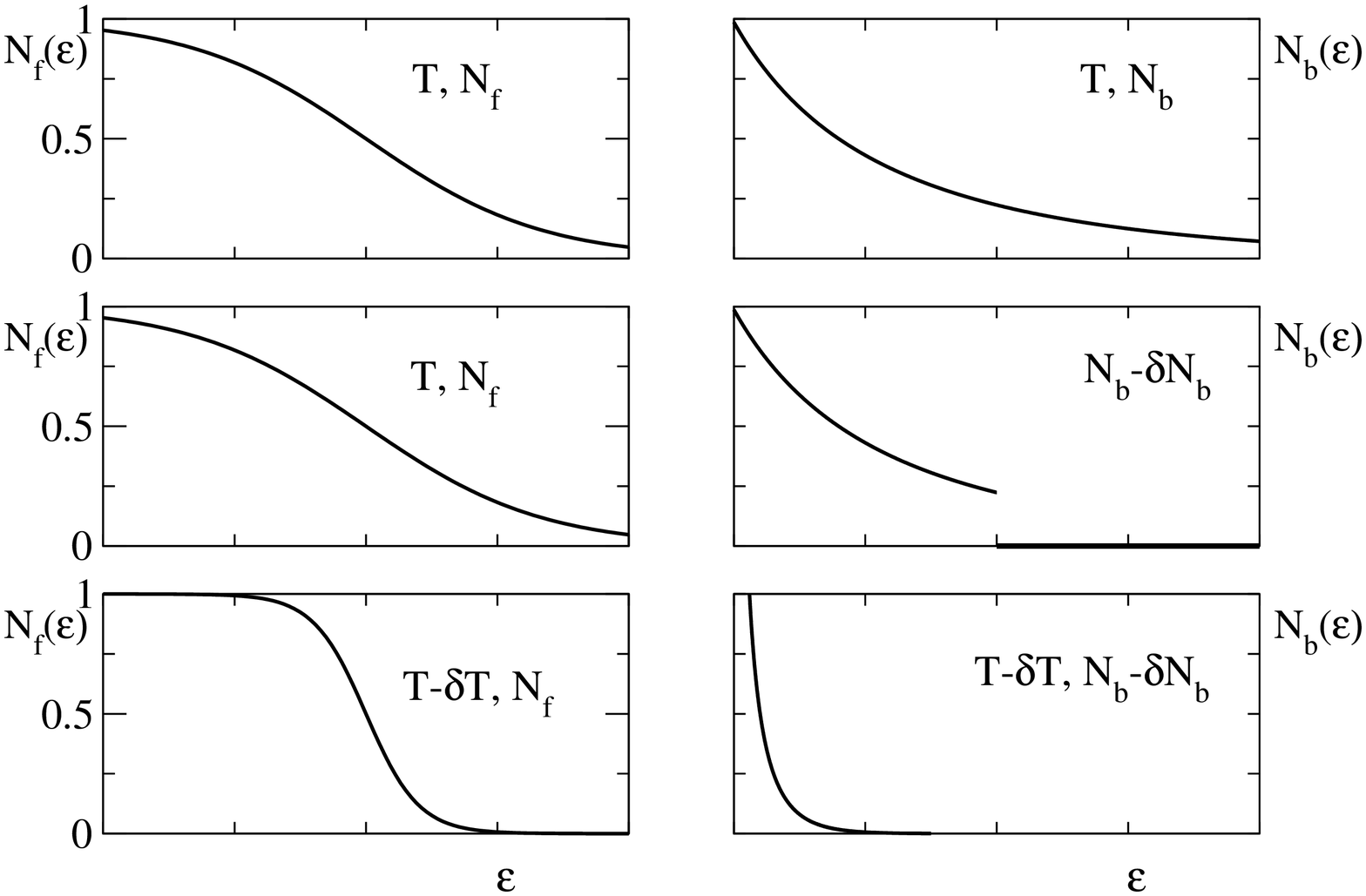}
\caption{Shown is a diagram of the evaporative cooling process in
our model. In the upper two panels, a gas of $N_f$ fermions at
temperature $T$ is in thermal equilibrium with a coolant of $N_b$
bosons or Boltzmann particles.  In the middle two panels, the tail of the
number distribution of the coolant is cut and $\delta N_b$
particles are removed. In the lower two panels, the systems are
permitted to rethermalize, resulting in a reduced common
temperature $T-\delta T$. The values of $\delta T$ and $\delta
N_b$ have been exaggerated for the purposes of illustration; the
units are arbitrary.\label{fig:1}}
\end{center}
\end{figure}
We assume that the Fermi gas, denoted by the subscript $f$, 
and the coolant, denoted by the subscript $b$, are initially
in thermal equilibrium at a common temperature $T$.  One
evaporates all particles of the coolant with an energy greater
than $\eta k_B T$~\cite{Mewes,hess1986,luiten1996,cct1997}, where $\eta$ is a constant, typically 5 to 7 in
current experiments.  This has the effect of removing a number of
particles $\delta N_b$ and an energy $\delta E_b$ from the
coolant:  \bea \delta N_b=\int_{\eta k_B T}^{\infty}
d\epsilon \,\rho_b(\epsilon) N_b(\epsilon)\, , \label{eqn:deltaN}\\
\delta E_b=\int_{\eta k_B T}^{\infty} d\epsilon \,\rho_b(\epsilon)
N_b(\epsilon) \epsilon\, ,\label{eqn:deltaE}\eea where
$\rho_b(\epsilon)$ is the density of states of a particle
in $b$ and $N_b(\epsilon)$ is the
mean occupation number of a state of energy $\epsilon$. 
In what follows we restrict ourselves to
the semi-classical regime in which the number of trap levels with
an energy less than $k_B T$ is much larger than unity, both for
species  $f$ and $b$. Then the case of a density of states of the
form \be \rho_{f,b}(\epsilon)=A_{f,b} \epsilon^{\delta+1/2}\ee may be
considered: this encompasses both a harmonic potential
$(\delta=3/2)$ and a box $(\delta=0)$~\cite{dos_comment}.
Here the subscript  $f$ ($b$) stands for the fermions (coolant).
In the two limiting
cases of a non-degenerate gas
$N_b(\epsilon)=\exp[-\beta(\epsilon-\mu)]$ and a Bose gas below $T_c$,
$N_b(\epsilon)=1/[\exp(\beta\epsilon)-1]$,  the mean energy of an
evaporated particle~\cite{cct1997} is independent of the chemical 
potential $\mu$ and proportional to $T$:
\be 
\frac{\delta E_b}{\delta N_b}=f(\eta)k_B T\,.
\ee
The function $f(\eta)$ is different for a Bose gas below $T_c$ 
and for a Boltzmann gas, but has the same  large $\eta$ behaviour
in both cases~\cite{formules}:
\be
f(\eta)= \eta+1+\frac{\delta+\frac{1}{2}}{\eta}
+\mathcal{O}\left(\frac{1}{\eta^2}\right)\, . \ee 

Once the particles are removed, the system is left to
rethermalize. The decrease in temperature $\delta T$ is obtained
by conservation of energy, 
\bea E_f(N_f,T)+E_b(N_b,T)-\delta E_b
\nonumber\\ = E_f(N_f,T-\delta T)+E_b(N_b-\delta N_b,T-\delta T)\,
.\label{eqn:cons}\eea  
Note that $E_f$ is the total energy of the fermions, not the Fermi energy.
The parameter $\eta$ is chosen sufficiently
large so that $\delta N_b \ll N_b$ and $\delta T \ll T$.
Equation~(\ref{eqn:cons}) may be expanded to leading order in
$\delta N_b$ and $\delta T$: \be \frac{\delta T}{\delta N_b}=
\frac{f(\eta)\,k_B T-\displaystyle\frac{\partial E_b}{\partial
N_b} (N_b,T)}{C_f(N_f,T)+C_b(N_b,T)}\, , \label{eqn:diffeq}\ee
where the heat capacities are defined by $C_f\equiv\partial_T E_f$ 
and $C_b = \partial_T E_b$.

Equation~(\ref{eqn:diffeq}), valid for a Boltzmann gas or 
a degenerate Bose
gas, will be studied for both a three dimensional box
and a harmonic potential. In the former case, the mean field
energy per particle  of the coolant, due to the interaction with the Fermi gas,
 is constant and therefore plays no role.  In the latter case, it  is not
exactly constant, but we shall assume that it is negligible
as compared to $k_B T$. Additionally, for simplicity it will be assumed that the
initial state of the fermions is degenerate.  In this case the
total energy is \bea E_f(N_f,T)=E_f(N_f,0)\nonumber\\ +\alpha N_f
k_B \frac{T^2}{T_F}\left\{1+\kappa \left(\frac{T}{T_F}\right)^2+
\mathcal{O}\left[\left(\frac{T}{T_F}\right)^4\right]\right\} \,
\label{eqn:T4}
,\eea where $\alpha\equiv \pi^2(\delta+3/2)/6$, $\kappa\equiv
\pi^2(\delta+1/2)(\delta-3)/20$,  $\delta$ is the exponent in
the density of states  and the Fermi temperature $T_F$ is
given by
\begin{equation}
A_f (k_B T_F)^{\delta+3/2} = (\delta+3/2) N_f.
\label{eqn:Af} \end{equation}
In Eq.~(\ref{eqn:T4}) we shall keep terms only to  second order in
$T$ in the following.

\section{Cooling by an ideal  Bose gas below $T_c$}
\label{sec:bose}

%%%%%%%%%%% figure 2 %%%%%%%%%%%
%
\begin{figure}[t]
\begin{center}
\epsfxsize=7.8cm \leavevmode \epsfbox{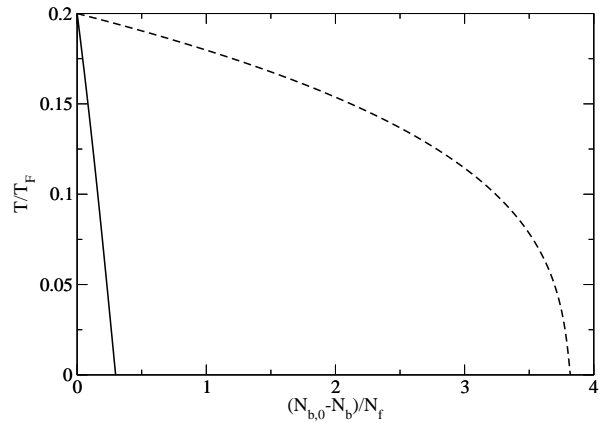}
\caption{Shown is the temperature of a degenerate Fermi gas
sympathetically cooled by a  Bose gas with a condensate present, 
as a function of the number of bosons evaporated in a harmonic trap, 
that is of the difference between the initial number of bosons $N_{b,0}$
(assumed to fulfill Eq.(\ref{eqn:cond_nat}))
and the final number $N_b$.  Two cases are
depicted:  $\omega_f/\omega_b=1$ (solid line) and
$\omega_f/\omega_b=5$ (dashed line) where $\omega_f/\omega_b$ is
the ratio of the mean trapping frequencies of the fermions and
bosons.  It is apparent that $\omega_f/\omega_b\lesssim 1$
requires a greatly reduced number of bosons in order to cool to
$T=0$, and may therefore be considered as a more efficient
experimental configuration.  Here the choice of parameters was
$\eta=6$ for the energy cut-off and $T_0/T_F=0.2$, 
where $T_0$ is the initial temperature. \label{fig:2}}
\end{center}
\end{figure}
A non-interacting  Bose gas with a temperature below $T_c$ 
has a total energy 
\be
E_b(N_b,T)=\frac{\nu A_b }{\delta+5/2}(k_B T)^{\delta+5/2}\, ,\ee
where $\nu\equiv (\delta+5/2)\Gamma(\delta+5/2)\zeta(\delta+5/2)$
is a constant and $\Gamma$ and $\zeta$ are gamma and zeta
functions, respectively~\cite{abramowitz1}.  Note that $E_b$ does
not depend on the total number of bosons, since the occupation of
the excited states is saturated to its maximal value in the
presence of a condensate.  Therefore $\partial E_b/\partial N_b =0$ and
Eq.~(\ref{eqn:diffeq}) reduces to \be
\frac{\delta T}{\delta N_b}=\frac{f(\eta)\,T_F}
{2\alpha N_f
\left[1+\displaystyle\frac{3\nu}{\pi^2}\frac{A_b}{A_f} 
\left(\frac{T}{T_F}\right)^{\delta+1/2}\right]}\,
\label{eqn:bosegen}\ee 
 where we used the fact that $A_f$ is related to the Fermi temperature $T_F$
by Eq.~(\ref{eqn:Af}).
When species $f$ and $b$ are in a box,  $A_b/A_f$ is simply the ratio
$V_b/V_f$ of the spatial volumes occupied by the two species.
When species $f$ and $b$ are trapped
harmonically, $A_b/A_f$ is given by the ratio $(\omega_f/\omega_b)^3$
where  $\omega_{f,b}$ are the geometric
means of the three oscillation frequencies for species  $f$ and
$b$, respectively.

The term in $T^{\delta+1/2}$ in the denominator of Eq.~(\ref{eqn:bosegen})
represents the ratio of the two heat capacities:  
\begin{equation}
\frac{C_b}{C_f }=
\frac{3\nu}{\pi^2}\frac{A_b}{A_f}
\left(\frac{T}{T_F}\right)^{\delta+1/2}.
\end{equation}
If one measures the efficiency of the cooling by the temperature
decrease per number of bosons evaporated, {\it i.e.}, $\delta T/ \delta N_b$,
it is most efficient to have $C_b\ll C_f$.  In this case the first
term in the denominator dominates and the derivative is maximized.
In Fig.~\ref{fig:2} is shown the temperature as a function of the
number of bosons evaporated, as can be obtained from  the solution
of the differential equation Eq.~(\ref{eqn:bosegen}) for the case of harmonic
trapping and $\eta=6$.  
It is apparent that it is advantageous to
evaporative cooling to choose a trap frequency ratio
$\omega_f/\omega_b \lesssim 1$, in order to maintain $C_b/C_f$
small at typical experimental temperatures $T/T_F \sim 0.2$ or
$0.1$. 

A simple condition may be
obtained from Eq.~(\ref{eqn:bosegen}) for the total initial number
of bosons necessary to cool to $T=0$: \be 
\label{eqn:cond_boson}
N_{b,0}>\Delta N_b = \frac{C_{f,0}}{k_B f(\eta)}\,
\left[1 +\frac{1}{\delta+3/2}\,\frac{C_{b,0}}{C_{f,0}} \right],\ee 
where $T_0$ is the initial temperature, $C_{0}$ are the initial heat capacities
and $\Delta N_b$ is the number of bosons evaporated. 
As the coolant is assumed to be initially  a Bose gas with a condensate present,
the initial number of bosons must  exceed the saturated value
of the population of the excited states:
\be
N_{b,0}  > A_b \,\Gamma(\delta+3/2)\,\zeta(\delta+3/2)\,(k_B T_0)^{\delta+3/2}.
\label{eqn:cond_nat}
\ee
Note that, as an arbitrarily large number of particles 
can be stored in the condensate, 
no connection can be established between the number of particles and
the heat capacity of the coolant.  Thus condition~(\ref{eqn:cond_boson})
cannot be considered as a constraint on $C_{b,0}/C_{f,0}$~\cite{implic}.

Recall that $\delta T/T \ll 1$ was required to replace
the finite differences in (\ref{eqn:cons}) by derivatives.
When the heat capacity of the  Bose gas is initially
much smaller that that of the Fermi gas, one finds that 
 this condition is satisfied for any value of $\eta$, including 
$\eta=0^+$  where $f(0)$ is finite and strictly positive.
In this case, Eq.~(\ref{eqn:bosegen}) and its consequences 
apply also when one evaporates all the non-condensed atoms at each cooling cycle.

\section{Cooling by a Boltzmann gas}
\label{sec:boltz}

A Boltzmann gas has a total energy \be E_b(N_b,T)=(\delta +3/2)N_b
k_B T \, .\ee  Equation~(\ref{eqn:diffeq}) then reduces to \be
\frac{\delta T}{\delta N_b}=\frac{[f(\eta)-(\delta+3/2)]T}{\displaystyle
2\alpha N_f \frac{T}{T_F}+(\delta+3/2)N_b}\, .\label{eqn:boltz}\ee
The explicit dependence of the energy on $N_b$ introduces a
negative contribution in the numerator.  However, 
the total numerator remains positive, since one can show, by integration
by parts, that the condition
\be f(\eta)>\delta+3/2\ee is satisfied for all $\eta>0$.

One finds that, by making a change of variables \be x\equiv
\frac{2\alpha}{\delta+3/2}\frac{T}{T_F}\frac{N_f}{N_b}=\frac{C_f}{C_b}\, ,\ee the
equation becomes separable. This leads to the constant of motion
\be
K=-\frac{\lambda}{\lambda-1}\ln|\lambda-1-x|+\frac{1}{\lambda-1}\ln
x -\ln (N_b/N_f)\, ,\ee where \be\lambda \equiv
\frac{f(\eta)-(\delta+3/2)}{\delta+3/2} > 0\, .\ee  The asymptotic
behavior of the temperature in the limit of the complete
evaporation of the Boltzmann particles, {\it i.e.}, $N_b\rightarrow 0$, falls
into two regimes.  The first is given by $x_0>\lambda-1$, where
$x_0$ is the initial value of $x$.  In this case, the temperature
tends to a finite non-zero value due to an insufficient number of
Boltzmann particles.  The ratio of the final to initial temperature is \be
\frac{T_{\mathrm{final}}}{T_0}=
\left[1-\left(\lambda-1\right)\frac{C_{b,0}}{C_{f,0}}\right]^{\lambda/(\lambda-1)}\,
.\ee

%%%%%%%%%%% figure 3 %%%%%%%%%%%
%
\begin{figure}[t]
\begin{center}
\epsfxsize=7.8cm \leavevmode \epsfbox{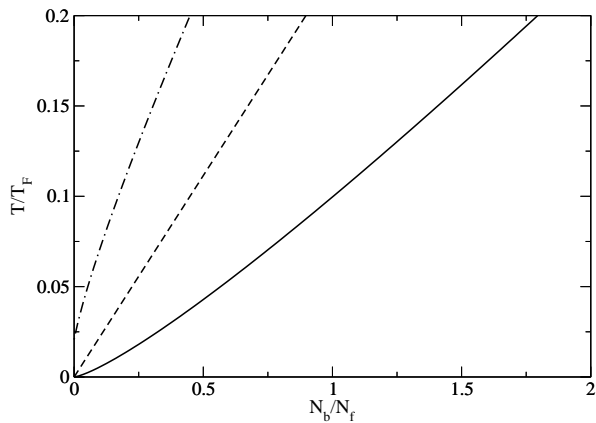}
\caption{Shown is the temperature of a degenerate Fermi gas
subject to sympathetic cooling by a Boltzmann gas as a function of
the number of Boltzmann particles. Three cases are depicted:
$C_{f,0}/C_{b,0}=(\lambda-1)/2$ (solid curve); $C_{f,0}/C_{b,0}=\lambda-1$ (dashed curve);
and $C_{f,0}/C_{b,0}=2(\lambda-1)$ (dot-dashed curve).  In the latter case the
number of Boltzmann particles is insufficient to cool to $T=0$. Here the
choice of parameters was $\eta=6$ for the energy cut-off parameter,
$T_0/T_F=0.2$ for the initial temperature, and a harmonic trapping
potential. \label{fig:3}}
\end{center}
\end{figure}

The second regime is given by $x_0<\lambda-1$, which implies
$\lambda>1$.  The  inequality $x_0<\lambda-1$
is  equivalent to the requirement that there be
a minimal initial number of Boltzmann particles:
\be
N_{b,0} > \frac{C_{f,0}}{k_B [f(\eta)-2(\delta+3/2)]}\, .
\label{eqn:require}\ee
Note that $\lambda>1$ implies that the denominator is positive.  
Equation~(\ref{eqn:require}) may be equally expressed as a requirement on
the initial heat capacities of the fermions and Boltzmann particles: \be
C_{f,0}<C_{b,0}(\lambda-1)\, .\ee In this case the temperature
tends to zero as \be \frac{T}{T_F}\propto
\left(\frac{N_b}{N_f}\right)^{\lambda}\, .\ee In Fig.~\ref{fig:3}
are shown three sample trajectories of Eq.~(\ref{eqn:boltz}). For
$x_0>\lambda-1$ the final temperature is non-zero when all of the
Boltzmann particles have been evaporated.  For $x_0<\lambda-1$ the final
temperature is zero and all of the Boltzmann particles are used in the
evaporation.  A third curve shows the case $x_0=\lambda-1$, which
divides the two regimes: this corresponds to constant $x$,
which implies $T/T_F\propto N_b/N_f$.

Note that, for a harmonic potential, $f(\eta)$ can be
calculated exactly for a Boltzmann gas~\cite{formules}.
The critical value of $\eta$ which corresponds to $\lambda-1=0$ is then
$\eta_c=4.59\ldots$. For $\eta<\eta_c$ the final temperature of
the fermions after evaporation of all the Boltzmann particles is always
non-zero.

\section{Conclusion}
\label{sec:conclusion}

We have demonstrated that, in an idealized experiment with no particle
losses besides those  due to evaporation of the coolant, sympathetic
cooling of degenerate Fermi gases is not limited by the heat capacity
of the coolant when the latter is a  Bose gas with a condensate
present.  In fact,
it is \emph{advantageous} to obtain $C_f/ C_b\gg 1$ in order to minimize the
number of bosons that must be evaporated, where $C_{f,b}$ are
the heat capacities of the fermions and bosons.  In
this case, $T=0$ may be obtained provided that the initial number
of bosons is sufficiently large: \be N_{b,0}>2.7 \,N_f T_0/T_F\,
,\ee where $N_f$ is the number of fermions, $T_0$ is the initial
temperature which is much less than the Fermi temperature $T_F$,
and $\eta=6$, a typical value of the energy cut-off for
evaporative cooling, was chosen.

In contrast, in the case of a Boltzmann coolant, the final
temperature of the fermions depends on the ratio of the initial values of
the heat capacities $C_{f,0}/C_{b,0}$.  For typical values of the evaporation cut-off
parameter $\eta$, $C_{f,0}/C_{b,0}$ must be smaller than a critical value
on the order of unity to reach $T=0$ after evaporation of all the Boltzmann particles.

{\bf Acknowledgments:} We would like to thank Thomas Bourdel and 
Jean Dalibard for useful discussions, and Zoran Hadzibabic and Roland
Combescot for useful comments on the manuscript. This work was
supported by NSF grant no. MPS-DRF 0104447. Laboratoire Kastler
Brossel is a research unit of l'Ecole Normale Sup\'erieure and
of l'Universit\'e Pierre et Marie Curie, associated with CNRS.  We
acknowledge financial support from R\'egion Ile de France.

%\bibliographystyle{prsty}
%\bibliography{refs}

\begin{thebibliography}{10}

\bibitem[*]{byline}
Present address: JILA, National Institute of Standards 
and Technology and Physics Department, University of Colorado,
Boulder, CO 80309-0440

\bibitem{truscott1}
A.~G. Truscott, K.~E. Strecker, W.~I. McAlexander, G.~Partridge, and R.~G. 
Hulet, Science {\bf 291},  2570  (2001).

\bibitem{schreck1}
F.~Schreck, L.~Khaykovich, K.~L. Corwin, G.~Ferrari, T.~Bourdel, J.~Cubizolles,
  and C.~Salomon, Phys. Rev. Lett. {\bf 87},  080403  (2001).

\bibitem{roati2002}
G. Roati, F. Riboli, G. Modugno, and M. Inguscio, Phys. Rev. Lett. {\bf 89},
  150403  (2002).

\bibitem{hadzibabic2002}
Z.~Hadzibabic, C.~A. Stan, K.~Dieckmann, S.~Gupta, M.~W. Zwierlein,
  A.~G\"orlitz, and W.~Ketterle, Phys. Rev. Lett. {\bf 88},  160401  (2002).

\bibitem{hadzibabic2003}
Z.~Hadzibabic, S.~Gupta, C.~A. Stan, C.~H. Schunck, M.W. Zwierlein,
  K.~Dieckmann, and W.~Ketterle, Phys. Rev. Lett.
{\bf 91}, 160401 (2003).

\bibitem{onofrio1}
R. Onofrio and C. Presilla, Phys. Rev. Lett. {\bf 89},  100401  (2002).

\bibitem{presilla2003}
C. Presilla and R. Onofrio, Phys. Rev. Lett. {\bf 90},  030404  (2003).

\bibitem{huang1}
K. Huang, {\em Statistical Mechanics}, 2nd ed. (John Wiley \& Sons, New York,
  NY, 1987).

\bibitem{constance} 
It is assumed for simplicity that the heat capacities are constant
in the range of temperatures considered.

\bibitem{hess1986}
H. F. Hess, Phys. Rev. B {\bf 34} 3476 (1986).

\bibitem{luiten1996}
O. J. Luiten, M. W. Reynolds, and J. T. M. Walraven, Phys. Rev. A {\bf 53} 381 (1996).

\bibitem{Mewes} K.B. Davis, M.O. Mewes, W. Ketterle, Appl. Phys. B {\bf 60}, 155 (1995).

\bibitem{wouters2002}
M. Wouters, J. Tempere, and J.~T. Devreese, Phys. Rev. A {\bf 66} 043414 (2002).

\bibitem{prudence} A gas of identical particles is in fact described by either
Bose or Fermi statistics as the temperature approaches zero.  However, for the
sake of highlighting the effect of Bose statistics in contrast to classical statistics
in the coolant, and in order to explain the meaning of the classical thermodynamic
argument mentioned above,
we shall imagine that Boltzmann statistics can be applied down to
zero temperature.  

\bibitem{proviso} We do not claim here that zero temperature can be 
achieved in a real experiment: it would require an infinite time,
as the collision rate between the fermions and the coolant tends
to zero for $T\rightarrow 0$. Furthermore the semi-classical approximation
used in this paper fails when the temperature drops below the energy of the
first excited state of a particle of the coolant.

\bibitem{cct1997}
C. Cohen-Tannoudji, Cours de physique atomique et mol\'eculaire, http://www.lkb.ens.fr/$\sim$cct,
  1997.

\bibitem{dos_comment}
Note that such a
density of states corresponds~\cite{cct1997} to a trapping
potential of form
$V(x,y,z)=a|x|^{1/\delta_x}+b|y|^{1/\delta_y}+c|z|^{1/\delta_z}$
where $\delta_x+\delta_y+\delta_z=\delta$.

\bibitem{formules} In the case of harmonic trapping, $\delta=3/2$ is
half integer so that $f(\eta)$ has a rather simple expression.
For a  Bose gas below $T_c$ we find
$f(\eta)=[\eta^3 g_1(e^{-\eta})+3\eta^2 g_2(e^{-\eta}) + 6\eta g_3(e^{-\eta})
+6 g_4(e^{-\eta})]/[\eta^2g_1(e^{-\eta}) + 2\eta g_2(e^{-\eta})
+2 g_3(e^{-\eta})]$ where the Bose function is $g_\alpha(z)=
\sum_{k\geq 1} z^k/k^\alpha$. For a Boltzmann gas
$f(\eta)=[\eta^3 +3\eta^2 + 6\eta +6]/[\eta^2+2\eta+2]$.

\bibitem{abramowitz1}
{\em Handbook of Mathematical Functions}, edited by M. Abramowitz and I.~A.
  Stegun (National Bureau of Standards, Washington, D. C., 1964).

\bibitem{implic}
One may wonder if the condition (\ref{eqn:cond_nat})
automatically implies (\ref{eqn:cond_boson}).
It turns out to be the case only if $\eta$ is large enough,
that is if 
$(\delta+5/2) C_{f,0} < \{[f(\eta)/f(0)]-[(\delta+5/2)/
(\delta+3/2)]\}C_{b,0}$.

\end{thebibliography}

\end{document}